\documentstyle[12pt,fullpage,psfig]{article}

\begin{document}

\hfill {\footnotesize ITP-SB-99-28, SUNY-NTG-99-23}\\
\vskip 1in
\begin{center} 
{\LARGE\bf 
On two-color QCD \\ with baryon chemical potential}
\vskip 1in
J.\,B. Kogut$^{\rm a}$, M.\,A. Stephanov$^{\rm b}$ and D. Toublan$^{\rm c}$
\\
\vskip 0.5in
{$^{\rm a}$ \it 
Loomis Laboratory of Physics, 1110 West Green Street, 
Urbana, IL 61801-3080, USA}
\\
{$^{\rm b}$ \it 
Institute for Theoretical Physics, SUNY, Stony Brook, NY 11794-3840, USA}
\\
{$^{\rm c}$ \it 
Department of Physics, SUNY, Stony Brook, NY 11794-3800, USA}
\vskip 1.5in

{\large\bf Abstract}
\end{center}

We study SU(2) color QCD with even number of quark flavors. First,
using QCD inequalities we show that at finite baryon chemical
potential $\mu$, condensation must occur in the channel with scalar
diquark quantum numbers. This breaks the U(1) symmetry generated by
baryon charge (baryon superconductivity). Then we derive the effective
Lagrangian describing low lying meson and baryon excitations using
extended {\em local} chiral symmetry of the theory. This
enables us to determine the leading term in the dependence of the
masses on $\mu$ exactly.

\newpage

\section{Introduction}

QCD at finite baryon number density has been intensely studied
recently~\cite{lattice.mu,St96b,SC,HaJa98,SRS}. 
Knowing the behavior of QCD in this regime will enable us
to understand the physics of heavy ion collisions, neutron
stars and supernova explosions. First principle calculations
using methods of lattice field theory have presented an insurmountable
theoretical challenge to date due to the absence of techniques to deal
numerically with complex measure path integrals. The two-color QCD model is an
exceptional case where conventional methods work due to
positivity of the Euclidean path integral measure~\cite{SU2lattice.old}. 

The two-color QCD model is also exceptional from the point of 
view of BCS-type diquark condensation
phenomenon which received attention recently~\cite{SC}.
In 3-color QCD such a condensate is not gauge invariant and it leads
to the phenomenon of color superconductivity. In 2-color QCD the
diquark condensate is a well-defined gauge invariant observable.
Before we learn how to deal with the three-color QCD it would be very
helpful to get as much insight as possible from the apparently 
easier (both conceptually and technically) case of two-color QCD.

Numerical calculations in SU(2) QCD are now being actively 
pursued~\cite{SU2lattice.new}.
In this letter we develop analytical methods which enable us 
to study two-color QCD at finite baryon number density, 
and in particular, to determine the spectrum of excitations.

We shall work in the Euclidean formulation of the theory. The Lagrangian
is given by:
\begin{equation}\label{action}
{\cal L} = \sum_{f=1}^{N_f}\left[\bar \psi_f \gamma_\mu D_\mu \psi_f
+ \mu\bar\psi_f\gamma_0\psi_f + m_q\bar\psi_f\psi_f\right], 
\end{equation}
and we shall omit the flavor indices $f$ in the following.
In this letter we shall consider the case of massless quarks, $m_q=0$.
The analysis of the more general massive case will be presented elsewhere.
We also illustrate our methods using the simplest case of
$N_f=2$ quark flavors.
The results can be easily extended to arbitrary even $N_f$. 
It is known that the 2-color $N_f=2$ theory with massless quarks 
at $\mu=0$ possesses SU(4)
global flavor symmetry which is broken spontaneously to Sp(4) \cite{SmVe}.
As a result the spectrum contains 5 Goldstone bosons. At finite $\mu$
the symmetry of the theory is reduced to the usual
SU(2)$\times$SU(2)$\times$U(1). We shall show, using QCD inequalities
and, independently, the exact effective Lagrangian, that this symmetry
is spontaneously broken down to SU(2)$\times$SU(2), creating a single
Goldstone boson corresponding to spontaneous breaking of baryon number
symmetry. The other 4 Goldstones acquire a common mass which is
proportional to $\mu$ for small $\mu\ll\Lambda_{\rm
QCD}$.%
\footnote{Such a linear dependence on $\mu$ can also be seen in the
simple effective sigma-model of Rapp et al.~\cite{SC}. Note that this
linear dependence of Goldstone masses on $\mu$ contrasts with 
the usual dependence on another
symmetry breaking parameter, the quark mass: $m_\pi\sim\sqrt{m_q}$.} 
We find that the coefficient of proportionality can be determined {\em
exactly}, and is equal to 2.

\section{QCD inequalities}

In Euclidean QCD, having a positive measure, one can
majorate all correlators with the correlator $\langle
\pi(x)\pi(0)\rangle$, where $\pi=\bar{u}\gamma_5d$ is the pion 
field~\cite{ineqs}.
Therefore, one can prove that $0^-$ is the lightest meson with
$I=1$. As a consequence, one obtains an important restriction on the
pattern of the symmetry breaking: it has to be driven by a condensate
$\langle\bar \psi\psi\rangle$ (not $\langle\bar \psi\gamma_5\psi\rangle$, for
example, which would give $0^+$ Goldstones).

Let us sketch the argument. Consider the Dirac operator in QCD: ${\cal
D} = \gamma\cdot (\partial + A) + \mu\gamma_0 + m_q$. When $\mu=0$
this operator obeys (matrix $A$ is antihermitian in Euclidean
formulation, while the $\gamma$-matrices are hermitian):
\begin{equation}
\gamma_5 {\cal D} \gamma_5 = {\cal D}^\dagger.	\label{5d5}
\end{equation}
Now consider the correlator of a generic meson: $M=\bar \psi\Gamma \psi$:
\begin{equation}
\langle M(x) M(0) \rangle
= \langle {\rm Tr} {\cal S}(x,0)\Gamma {\cal S}(0,x) \Gamma \rangle,
\end{equation}
where we did the obvious integration over the $\psi$'s and
$\bar\psi$'s and left the
integration over the $A$'s (it is important, as is true for $I=1$, 
that there is no disconnected piece). ${\cal S}\equiv {\cal D}^{-1}$. 
When $\Gamma=\gamma_5$
we can use (\ref{5d5}) to rewrite the expression in brackets as:
\begin{equation}
{\rm Tr} {\cal S}(x,0) {\cal S}^\dagger(x,0)
\end{equation}
which is manifestly positive (the dagger in this formula only transposes
the color and Dirac indices, the coordinate indices $x,0)$ we
transposed explicitly). Moreover, for any  $\Gamma$ (such that $\Gamma^2=1$) 
we can write, using the Schwartz inequality:
\begin{equation}
{\rm Tr} {\cal S}(x,0)\Gamma {\cal S}(0,x) \Gamma
= {\rm Tr} {\cal S}(x,0) \Gamma \gamma_5 {\cal S}^\dagger(x,0) \gamma_5\Gamma
\le {\rm Tr} {\cal S}(x,0) {\cal S}^\dagger(x,0).
\end{equation}
If the measure is positive
this inequality should survive the averaging, and we get the
desired inequality for the correlators, and therefore for the
meson masses.%
\footnote{To make this argument into a mathematical theorem one would
have to consider regularization. We shall not pursue this level
of rigor here and shall refer the interested reader to original
literature~\cite{ineqs}.}

For $\mu\ne0$ we lose the positivity and we lose the inequalities
in SU(3). But, in SU(2) QCD we also have a positive measure! Can we,
perhaps, derive some inequalities for the meson masses and
consequently make some conclusions about the symmetry breaking
pattern?

The relation (\ref{5d5}) holds in either SU(3) or SU(2). It also
fails in both theories at $\mu\ne0$. But there is another
relation, which holds in SU(2), due to its pseudo-reality, 
for {\em arbitrary\/} $\mu$:
\begin{equation}
\gamma_5 C T_2 {\cal D} \gamma_5 C T_2 = {\cal D}^*,
\end{equation}
where $C=i\gamma_0\gamma_2$ ($C^2=1$, $C\gamma_\mu C=-\gamma_\mu^*$) 
all $\gamma$-matrices are hermitian,
and $T_2$ is a generator of the SU(2) color (the second Pauli
matrix, $T_2T_aT_2=-T_a^*$). It is a consequence of this
relation that the measure is positive, in fact.

If we construct now the correlator of the diquark $M_{\psi\psi}=\psi^T C T_2
\gamma^5 \psi$ (this is $0^+$, $I=0$, i.e., antisymmetric in flavor),
we have:
\begin{equation}
\langle M_{\psi\psi}(x) M_{\psi\psi}^\dagger(0) \rangle
= \langle {\rm Tr} {\cal S}(x,0) C T_2 \gamma^5
{\cal S}^T(x,0) C T_2 \gamma^5 \rangle
= \langle {\rm Tr} {\cal S}(x,0) {\cal S}^\dagger(x,0) \rangle.
\end{equation}
Now, as before, one can show that the correlator of $\psi^T C T_2\gamma^5 \psi$
meson majorates a correlator of any other meson $\psi^T C T_2\gamma^5
\Gamma \psi$. In particular, we see that it
is $0^+$, not $0^-$, which is the lightest. 
Therefore, if there
is condensation it has to be that of $\psi^T C T_2\gamma^5 \psi$, not
violating parity, in particular. 

One can also majorate the correlator
of any meson of the type $\bar \psi\Gamma \psi$. In other words, the $0^+$
diquark must be the lightest meson in this case. This excludes
the possibility of conventional condensation of
 $\langle\bar\psi\psi\rangle$ which otherwise 
would lead to 3 massless pions (unless the inequality is saturated,
which is the case at $\mu=0$).

\section{Symmetries, breaking and Goldstone counting}

We shall construct the effective Lagrangian describing light
excitations in 2-color QCD at finite $\mu$ in the next section.
Here we shall analyze the global symmetries of our theory
--- a necessary ingredient of this construction.

We start from the known case of $\mu=0$ and recall the fact that
the global symmetry of the theory is SU($2N_f$) rather than the
usual SU($N_f$)$\times$SU($N_f$)$\times$U(1) \cite{SmVe}. This can 
be seen explicitly by using left and right chiral Weyl components
of the Dirac spinor $\psi=(q_L,q_R)$:
\begin{equation}\label{qaction}
{\cal L} = \bar \psi \gamma_\mu D_\mu \psi = q_L^\dagger i\sigma_\mu D_\mu q_L
+ q_R^\dagger i\bar\sigma_\mu D_\mu q_R, 
\end{equation}
where $\sigma_\mu=(-i,\sigma_k)$ and $\bar\sigma_\mu =
(-i,-\sigma_k)$, and $\sigma_k$ are usual Pauli matrices.
The fact that (\ref{action}) has higher flavor symmetry
is based on the property of the 2-color Dirac operator,
$D_\mu = \partial_\mu + A_\mu$ ($A$ is antihermitian SU(2) color
generator matrix $A=A^aT_a$, where $T_a$ are color generators):
\begin{equation}
D_\mu^T = - T_2 D_\mu T_2 
\end{equation}
which in turn is based on the (pseudoreality) property of the
generators of SU(2) (Pauli matrices):
\begin{equation}\label{pr}
T_a^* = (T_a^T = )  - T_2T_aT_2.
\end{equation}
We introduce:
\begin{equation}\label{tilde}
\tilde q = \sigma_2 T_2 q_R^\dagger, \quad \mbox{and}\quad 
\tilde q^\dagger = q_R^T T_2 \sigma_2.
\end{equation}
We then substitute (\ref{tilde}) into (\ref{qaction})
and use the property (\ref{pr}) of Pauli matrices for both 
$T_a$ of color and $\sigma_k$ of Euclid, together with the
anticommutativity of $\tilde q$, $\tilde q^\dagger$ (we need
to transpose) to arrive at:
\begin{equation}
{\cal L} = q^\dagger i\sigma_\mu D_\mu q +
\tilde q^\dagger i\sigma_\mu D_\mu \tilde q =
\Psi^\dagger i\sigma_\mu D_\mu \Psi.
\end{equation}
which now has a manifest SU($2N_f$) ``flavor'' symmetry. The $\Psi$ denotes
a Weyl spinor which has $2N_f$ ``flavor'' components. E.g., for $N_f=2$:
\begin{equation}
\Psi = \left(\begin{array}{c}q\\ \tilde q\end{array}\right)
=\left(\begin{array}{c}q^1\\q^2\\ \tilde q^1\\ \tilde q^2\end{array}\right),
\end{equation}
where $1,2$ are the original flavor indices.

The total global symmetry of the action is SU($2N_f$)$\times$U(1)$_A$. Note
that the baryon symmetry, under which $B(q)=+1$ and $B(\tilde q)=-1$
is a subgroup of this SU($2N_f$). The $\tilde q$ are, therefore,
conjugate quarks (since they have opposite baryon charge to normal
quarks $q$) in the terminology of \cite{St96b}. 
Under axial U(1)$_A$ $q$ and $\tilde q$ have the
same charge (because $A(\tilde q) = -A(q_R) = A(q_L) = A(q)$). This
symmetry is broken by the anomaly, however, so the actual symmetry
of the quantum field theory is SU($2N_f$).

Now, let us write down various useful quark bilinears in terms of
$q$, $\tilde q$ and determine
their transformation properties under this SU($2N_f$).
\begin{equation}\label{pbp}
\bar\psi\psi = q_R^\dagger q_L + q_L^\dagger q_R
=
\tilde q^T \sigma_2 T_2 q 
+ q^\dagger \sigma_2 T_2 (\tilde q^\dagger)^T 
=
\frac12 \Psi^T \sigma_2 T_2 
\left(\begin{array}{cc}0&1\\-1&0\end{array}\right) \Psi
+ \mbox{h.c.}\ .
\end{equation}
The matrix in (\ref{pbp}) is a $2N_f\times2N_f$ matrix in the SU($2N_f$)
indices. The matrices $\sigma_2$ and $T_2$ carry SU(2) spin
and SU(2) color indices respectively and no SU($2N_f$) indices.
They are just antisymmetric $\epsilon$-symbols for their indices.
We see that the chiral condensate is not a singlet under SU($2N_f$).
Since it is an antisymmetric product of two fundamental SU($2N_f$) spinors
$\Psi$, it transforms as an antisymmetric tensor of rank 2. The dimension
of this representation is $N_f(2N_f-1)$.

We shall continue our discussion using $N_f=2$ case as an example.
For $N_f=2$ (\ref{pbp}) transforms as a 6-plet.
The (\ref{pbp}) gives us one component of this 6-plet (sigma). The
remaining 5 are: 3 pions, scalar diquark and anti-diquark.

What does the chemical potential do? 
\begin{equation}\label{pb0p}
\bar\psi\gamma_0\psi = q_L^\dagger q_L + q_R^\dagger q_R
= q^\dagger q + \tilde q^T (\tilde q^\dagger)^T =
q^\dagger q - \tilde q^\dagger \tilde q
= \Psi^\dagger\left(\begin{array}{cc}1&0\\0&-1\end{array}\right)\Psi.
\end{equation}
We see that this term is not a singlet under SU(4).
Since $4\times4=1+15$ it is a component of a 15-plet (adjoint
representation $(2N_f)^2-1$).
It is easy to understand the meaning of $+1$ and $-1$ in the matrix
in (\ref{pb0p}) ---
these are just baryon charges of quarks and conjugate quarks.

What is the remaining subgroup of SU(4), under which (\ref{pb0p}) is
invariant? From the block-diagonal structure of (\ref{pb0p})
it is clear that SU(2)$_L\times$SU(2)$_R$ rotations preserve it, since
these rotate the first two components of $\Psi$, or the last two,
separately. The U(1)$_B$, which can be thought as generated by the block
$\tau_3$ generator (the charges are $(+,+,-,-)$),
also preserves (\ref{pb0p}). All other generators are broken by (\ref{pb0p}).
To summarize, we start with an SU(4) symmetry; then
we add a term proportional to $\mu$ which transforms as a component of a
15-plet of this SU(4) which
breaks this symmetry {\em explicitly} down to SU(2)$_L\times$SU(2)$_R\times$
U(1)$_B$. 

Now let us do the Goldstone counting.
At $\mu=0$ we have SU(4) global symmetry. The non-zero expectation value of
the quark bilinear (\ref{pbp}) which develops spontaneously
breaks it down to Sp(4). This produces 5 Goldstone bosons (15
generators minus 10).
On the other hand, when $\mu\ne0$ the symmetry of the theory
is SU(2)$_L\times$SU(2)$_R\times$U(1)$_B$. As we concluded in
the previous section this symmetry should break down to
SU(2)$_L\times$SU(2)$_R$ by the non-zero expectation value of
scalar diquark. Therefore at $\mu\ne0$ the theory has only
one Goldstone. What happened to the other 4? As is easy
to guess, and as we shall see explicitly, they form a representation
$(\underline2,\underline2)$ of the manifest SU(2)$_L\times$SU(2)$_R$
group, and acquire the same mass. This mass should vanish at $\mu=0$.
In the next section we shall calculate
the dependence of the mass of the 4-plet of these pseudo-Goldstones,
$m_{\rm pG}$ as a function of $\mu$ for small $\mu$. 

\section{Effective Lagrangian}

\subsection{Global symmetry}

In this section we construct the effective Lagrangian for the low
energy degrees of freedom, which in our theory with spontaneous
symmetry breaking are the Goldstone bosons~\cite{ChiL}.
The basic steps we follow are:
(i) identify the symmetries of the underlying (microscopic) theory; 
(ii) identify degrees of freedom of the effective (macroscopic) theory;
(iii) ensure that the effective theory is invariant
under the symmetries of the microscopic theory.%
The microscopic theory at $\mu=0$ has a global
SU(4) symmetry. In the effective theory, which we want to construct,
the degrees of freedom are given by  the fluctuations of the
condensate of $\Sigma$:
\begin{equation}
\Sigma \sim \Psi\Psi^T\sigma_2T_2,
\end{equation}
which is a Lorentz and color singlet but flavor SU(4) 6-plet.
Fluctuations of the orientation of $\Sigma$ give us our 5 Goldstones.
Under the action of $U\in$SU(4):
\begin{equation}\label{psitran}
\Psi\to U\Psi 
\end{equation} 
and thus
\begin{equation}\label{sigmatran}
\Sigma\to U\Sigma U^T.
\end{equation}  
The low-energy effective Lagrangian 
invariant under the flavor SU(4) can be written as a non-linear sigma
model~\cite{ChiL,SmVe}:
\begin{equation}\label{lagr1}
{\cal L}_{1} = 
f_\pi^2 {\rm Tr }\partial_\mu \Sigma^{\dagger} \partial_\mu \Sigma \ .
\end{equation}
The matrix $\Sigma$ in the effective Lagrangian is a unitary
antisymmetric matrix (which has exactly 5 independent real parameters).
The degrees of freedom are the rotations of $\Sigma$ generated by
$U$ as in (\ref{sigmatran}). The transformations $U$ which leave $\Sigma$
invariant form the Sp(4) group. The nontrivial degrees of freedom of the
Lagrangian (\ref{lagr1}) -- the Goldstones -- live in the coset SU(4)/Sp(4).

In the microscopic theory, the term:
\begin{equation}\label{muterm}
\mu\bar\psi\gamma_0\psi = 
\mu\Psi^\dagger\left(\begin{array}{cc}1&0\\0&-1\end{array}\right)\Psi.
\end{equation}
breaks the SU(4) symmetry explicitly. 
However, we can save this symmetry by transforming also the source
coupled to the breaking term. Rewriting (\ref{muterm}) as
\begin{equation}
\mu\Psi^\dagger i\sigma_\mu B_\mu\Psi,
\end{equation}
where $B_\mu$ is an SU(4) matrix. The value of $B_\mu$
fixed by (\ref{muterm}) is:
\begin{equation}\label{bfixed}
B_\mu = \delta_{0\mu}\left(\begin{array}{cc}1&0\\0&-1\end{array}\right)
\end{equation}
If under the
transformation (\ref{psitran}) the matrix $B_\mu$ also transforms as:
\begin{equation}\label{btran}
B_\mu \to UB_\mu U^\dagger,
\end{equation}
the microscopic Lagrangian
\begin{equation}
\Psi^\dagger i\sigma_\mu D_\mu \Psi + 
\mu\Psi^\dagger i\sigma_\mu B_\mu\Psi
\end{equation}
will be invariant. Thus the effective
Lagrangian must be also invariant under such an extended
transformation. For example, this requirement rules out the term
linear in $B$ (and therefore in $\mu$) in the effective Lagrangian. 
This is because $B$ transforms as (\ref{btran}) under SU(4), and one cannot
construct a non-trivial invariant out of $\Sigma$ and only one power of $B$.
The lowest order nontrivial term which we can write is:
\begin{equation}\label{aterm}
\mu^2{\rm Tr } \Sigma B^T \Sigma^{\dagger} B.
\end{equation}
This term will produce the mass for the Goldstone bosons {\em linear} in
$\mu$. 

\subsection{Local symmetry}

We see that the symmetry considerations help us find the form of the
symmetry breaking term in the effective Lagrangian, and thus determine
the dependence of
the $m_{\rm pG}$ on $\mu$. However, it does not tell us what the 
coefficient of proportionality in $m_{\rm pG}\sim \mu$ is, since
it does not specify the coupling of (\ref{aterm}). 
This coefficient can be determined
if we notice that the global symmetry (\ref{psitran}), (\ref{btran})
can in fact be promoted to a {\em local} symmetry in the microscopic
theory~\cite{ChiL}. This will require the transformation of $B$:
\begin{equation}\label{btranlocal}
B_\mu \to UB_\mu U^\dagger + {1\over\mu}U\partial_\mu U^\dagger.
\end{equation}
In order to ensure that the effective Lagrangian is also invariant
under this local symmetry we have to replace the derivatives in
(\ref{lagr1}) by the long covariant derivatives:
\begin{eqnarray}
D_\mu \Sigma &=& \partial_\mu \Sigma + \mu(B_\mu\Sigma + \Sigma B_\mu^T)
\nonumber\\
D_\mu \Sigma^{\dagger} &=& \partial_\mu \Sigma^{\dagger} -
\mu(\Sigma^{\dagger}B_\mu + B_\mu^T\Sigma^{\dagger}).
\end{eqnarray}
The signs here are important and are fixed by the local symmetry. 
The Lagrangian must have the form:
\begin{equation}\label{lagr2}
{\cal L}_{\rm eff} = f_\pi^2 {\rm Tr } D_\mu \Sigma^{\dagger} D_\mu \Sigma
\end{equation}
Expanding the long derivatives we find:
\begin{equation}\label{lagr3}
{\cal L}_{\rm eff} = f_\pi^2 \left[ 
{\rm Tr } \partial_\mu \Sigma^{\dagger} \partial_\mu \Sigma
- 4\mu{\rm Tr } \partial_\mu \Sigma \Sigma^{\dagger} B_\mu
- \mu^2{\rm Tr }(\Sigma^{\dagger}B_\mu + B_\mu^T\Sigma^{\dagger})
(B_\mu\Sigma + \Sigma B_\mu^T)
\right] \ ,
\end{equation}
where we used the property $\Sigma^T=-\Sigma$ to simplify the second term.
Now we shall analyze the last term in (\ref{lagr3}).
There are two main effects of this term. First, the minimum of it
with respect to all possible orientations of $\Sigma$ obtained by
rotations (\ref{sigmatran}) determines the direction of the
condensation. Second, the curvature matrix around this minimum
gives us the masses for the (pseudo)-Goldstones.

We see that the local symmetry relates the mass term to the kinetic
term in the Lagrangian. This means that the coefficient of
proportionality in the equation $m_{\rm pG}={\rm const}\cdot\mu$, which
is just a {\em dimensionless} number, is fixed by the local chiral
symmetry! In
particular, $f_\pi$ does not enter at all into this relation.
In the remainder of this note we shall set $f_\pi=1$ to simplify
the formulas.

\subsection{Vacuum alignment}

Using the fact that
$B_\mu^\dagger=B_\mu$, we can see that the last term in (\ref{lagr3})
is seminegative definite:
\begin{equation}\label{3rdterm}
{\cal L}_{3} = - \mu^2{\rm Tr } AA^\dagger,
\end{equation}
where
\begin{equation}
A = B\Sigma + \Sigma B^T.
\end{equation}
In order to find the vacuum alignment of $\Sigma$ we must minimize
${\cal L}_3$. 
Let us try first the alignment corresponding to the usual
chiral condensate:
\begin{equation}
\Sigma = \Sigma_{\bar\psi\psi} \equiv 
\left(\begin{array}{cc}0&-1\\1&0\end{array}\right).
\end{equation}
Using $B$ given by (\ref{bfixed}) we find that $A=0$ and therefore
${\cal L}_3=0$ which is the absolute maximum of ${\cal L}_3$, not the minimum
which we seek. Therefore the standard vacuum alignment (with no baryon
charge in the condensate) is unstable. 

One can see that the minimum can be achieved for a
$\Sigma=\Sigma_0$ such that:
\begin{equation}\label{bsigma}
B\Sigma_0 = \Sigma_0 B^T.
\end{equation}
A solution to (\ref{bsigma}) is given by:
\begin{equation}\label{sigma0}
\Sigma_0 = \left(\begin{array}{cc}\sigma_2&0\\0&\sigma_2\end{array}\right).
\end{equation}
This minimum is not unique, there is a U(1) degeneracy, corresponding
to the rotation with the generator given by $B_0$ (\ref{bfixed}).
This gives the Goldstone corresponding to the spontaneous 
breaking of the baryon charge symmetry.
Any other rotation will raise the value of the effective potential.

\subsection{Mass spectrum}

Now we shall consider the curvature of the potential in more detail,
to determine $m_{\rm pG}$. We can rewrite (\ref{3rdterm}) as:
\begin{equation}\label{l3}
{\cal L}_{3} = 
-2\mu^2{\rm Tr } B_\mu^2 
-2\mu^2{\rm Tr }\Sigma B_\mu^T \Sigma^{\dagger} B_\mu.
\end{equation}
The dependence on $\Sigma$ sits in the second term. In order to find
the mass matrix for the (pseudo)-Goldstones we should expand
$\Sigma$ in small fluctuations around the vacuum value $\Sigma_0$
(\ref{sigma0}). These small fluctuations are given in terms of the
transformation (\ref{sigmatran}) with $U$ close to unity:
\begin{equation}\label{sigmasigma0}
\Sigma = U \Sigma_0 U^T.
\end{equation}
We shall write $U$ as an exponent of the generators of the SU(4).
But first let us, following the formalism and notations of 
Peskin~\cite{peskin},
separate the generators into those which do not change $\Sigma_0$ ---
$T_i$, and those that do --- $X_a$. The transformations $U$ generated
by $T_i$:
\begin{equation}\label{Ti}
U = e^{i\phi_i T_i} \qquad \mbox{such that} \qquad U\Sigma_0U^T=\Sigma_0,
\end{equation}
form an Sp(4) subgroup of SU(4). It follows from (\ref{Ti})
that these generators obey:
\begin{equation}
T_i\Sigma_0 = -\Sigma_0 T_i^T.
\end{equation}
The remaining 5 generators $X_a$ can be shown, using the block
representation of Peskin, to obey:
\begin{equation}\label{xsigma}
X_a\Sigma_0 = +\Sigma_0 X_a^T.
\end{equation}
The corresponding fields $\pi_a$ defined as:
\begin{equation}\label{ux}
U = e^{i\pi_a X_a},
\end{equation}
and by (\ref{sigmasigma0}) 
are the dynamical degrees of freedom of the Lagrangian (\ref{lagr3}).

We shall write here, to provide an example, the explicit form of the
generators $T$ and $X$ in our case of the SU(4) flavor group. There are
10 generators $T_a$:
\begin{equation}
T_{1-3}=\left(\begin{array}{cc}\sigma_i&0\\0&\sigma_i\end{array}\right),
\ 
T_{4-6}=\left(\begin{array}{cc}\sigma_i&0\\0&-\sigma_i\end{array}\right),
\ 
T_{7-9}=\left(\begin{array}{cc}0&\sigma_i\\\sigma_i&0\end{array}\right),
\ 
T_{10}=\left(\begin{array}{cc}0&i\\-i&0\end{array}\right).
\end{equation}
And here are the 5 generators $X_a$:
\begin{equation}
X_{1-3}=\left(\begin{array}{cc}0&i\sigma_i\\-i\sigma_i&0\end{array}\right),
\qquad
X_4=\left(\begin{array}{cc}0&1\\1&0\end{array}\right),
\qquad
X_5=\left(\begin{array}{cc}1&0\\0&-1\end{array}\right).
\end{equation}
We have normalized the generators as:
\begin{equation}\label{tx}
{\rm Tr } T_i T_k = \delta_{ik}{\rm Tr } 1,
\quad
{\rm Tr } X_a X_b = \delta_{ab}{\rm Tr } 1,
\quad
{\rm Tr } X_a T_i = 0,
\end{equation}
where ${\rm Tr } 1=4$.

Let us make the following observations. First,
$B$ should be one of the generators
of SU(4). From (\ref{bsigma}) and (\ref{xsigma}) we conclude
that it has to belong to the set of broken generators $X_a$. Our
explicit example confirms this, indeed $X_5=B$.
Second, the remaining generators ($a=1,2,3,4$ in our example) 
anticommute with $B$.

For a transformation $U$ generated by $X$ which {\em commutes} with $B$
we can write for the last term in (\ref{l3}):
\begin{equation}
- 2{\rm Tr }\Sigma B_\mu^T \Sigma^{\dagger} B_\mu = 
- 2{\rm Tr }U\Sigma_0 U^T B_\mu^T U^*\Sigma_0^{\dagger}U^{\dagger} B_\mu =
- 2{\rm Tr } B_\mu^2,
\end{equation}
which is a constant. So there is no mass for the corresponding
boson. We conclude that $\pi_5$ is a true Goldstone.

For the remaining 4 generators $X$ we can write, using the fact that
they anticommute with $B$:
\begin{eqnarray}
- 2{\rm Tr }\Sigma B_\mu^T \Sigma^{\dagger} B_\mu 
&=& 
- 2{\rm Tr }U\Sigma_0 U^T B_\mu^T U^*\Sigma_0^{\dagger}U^{\dagger} B_\mu
\nonumber \\
&=& - 2{\rm Tr } U^2\Sigma_0 (U^T)^2 B_\mu^T\Sigma_0^{\dagger} B_\mu 
= - 2{\rm Tr } U^4,
\end{eqnarray}
where we have used properties of the $X$ generators (\ref{xsigma}), 
(\ref{bsigma}) and $B^2=1$. Now using (\ref{ux}), (\ref{tx}) and expanding
to quadratic order in the fields we find:
\begin{equation}\label{massterm}
- 2{\rm Tr } U^4 = + 16\pi_a\pi_b{\rm Tr } X_a X_b + {\cal O}(\pi^4)
= 16\pi_a^2 {\rm Tr } 1 + {\cal O}(\pi^4),
\end{equation}
where $a=1-4$.
Now expanding the kinetic term we find, using (\ref{sigmasigma0})
and (\ref{xsigma}):
\begin{equation}\label{kinterm}
{\rm Tr } \partial_\mu \Sigma^{\dagger} \partial_\mu \Sigma
= {\rm Tr } \partial_\mu (U^{-2}) \partial_\mu (U^2) 
= 4 \partial_\mu\pi_a\partial_\mu\pi_b {\rm Tr } X_a X_b + {\cal O}(\pi^4)
= 4 (\partial_\mu\pi_a)^2 {\rm Tr } 1+ {\cal O}(\pi^4),
\end{equation}
where $a=1-5$. Desired normalization of the kinetic term
can be trivially achieved by rescaling the fields $\pi_a$. 
Taking together (\ref{massterm}), (\ref{kinterm})
and (\ref{lagr3}) we find:
\begin{equation}\label{mpg}
m_{\rm pG} = m_{1-4} = 2\mu \qquad \mbox{and} \qquad m_5 = 0.
\end{equation}
This is our result.

\subsection{Linear term}

What is the significance of the second term in (\ref{lagr3}):
\begin{equation}\label{2term}
- 4\mu{\rm Tr } \partial_\mu \Sigma \Sigma^{\dagger} B_\mu\ ?
\end{equation}
Let us expand it for the generators $X$ using (\ref{tx}):
\begin{equation}
- 4\mu{\rm Tr } \partial_\mu \Sigma \Sigma^{\dagger} B_\mu
= -8i\mu \partial_\mu \pi_5{\rm Tr } X_5 B_\mu + {\cal O}(\pi_5^3).
\end{equation}
Only the generator $X_5$ gives a nonvanishing contribution to the linear
term in $\pi$, and the fact that ${\rm Tr }X_5X_5B=0$ ensures that
there are no terms quadratic in $\pi_5$.  This linear term means that
$B_\mu$ (the baryon charge current) is a source of the Goldstone field
$\pi_5$ (similar to the axial current in QCD being the pion source).

\section{Generic even $N_f$}

Most of the derivation goes through {\it mutatis mutandis} in the
general case. Here we shall summarize the results.
The global symmetry at $\mu=0$ is SU($2N_f$).
This symmetry is broken spontaneously:
\begin{equation}
{\rm  SU}(2N_f)\stackrel{\Sigma}{\to} {\rm  Sp}(2N_f), \qquad (\mu=0).
\end{equation}
The number of the Goldstones in this case is:
\begin{equation}\label{gmu=0}
((2N_f)^2-1) - \frac12\ 2N_f(2N_f+1) = 2N_f^2 -N_f -1, \qquad (\mu=0).
\end{equation}
At nonzero $\mu$ the symmetry is broken down to:
\begin{equation}
{\rm  SU}(2N_f)\stackrel{\mu}{\to}{\rm  SU}(N_f)\times {\rm  SU}(N_f)\times {\rm  U}(1).
\end{equation}
The condensate, being an antisymmetric rank 2 tensor (cf. (\ref{sigma0}),  
breaks it now in the following way:
\begin{equation}\label{break1}
{\rm  SU}(N_f)\times {\rm  SU}(N_f)\times {\rm  U}(1)\stackrel{\Sigma}{\to}
{\rm  Sp}(N_f)\times {\rm  Sp}(N_f).
\end{equation}
Note that in the case $N_f=2$: Sp(2)$\sim$ SU(2), so
only U(1) is broken. When $N_f>2$ we shall have
more than one true Goldstone boson. Their number is:
\begin{equation}\label{gmune0}
2(N_f^2-1) + 1 - 2\frac12 N_f(N_f+1) = N_f^2 - N_f - 1 \qquad (\mu\ne0).
\end{equation}

Comparing (\ref{gmu=0}) and (\ref{gmune0}) we find that there are
$N_f^2$ pseudo-Goldstone bosons. Their masses are given by
(\ref{mpg}): $m_{\rm pG}=2\mu$ for small $\mu$.

In terms of group representations, we have the following picture. First,
$\mu=0$.  The fermions transform as a fundamental $2N_f$-plet under
SU($2N_f$). The fermion condensate transforms as an antisymmetric tensor
of rank 2. The dimension of this representation is $N_f(2N_f-1)$.
After the breaking to Sp($2N_f$) the
Goldstones fall into an irreducible representation of Sp($2N_f$)
given by the antisymmetric tensor of rank
2 with the condition that trace of that tensor times the matrix $\Sigma_0$
is zero. The dimension of this representation is $N_f(2N_f-1)-1$
which is exactly (\ref{gmu=0}).

The baryon charge current to which $\mu$ couples transforms in the
adjoint representation of SU($2N_f$), which has dimension
$(2N_f)^2-1$. 

\begin{figure}
\centerline{\psfig{file=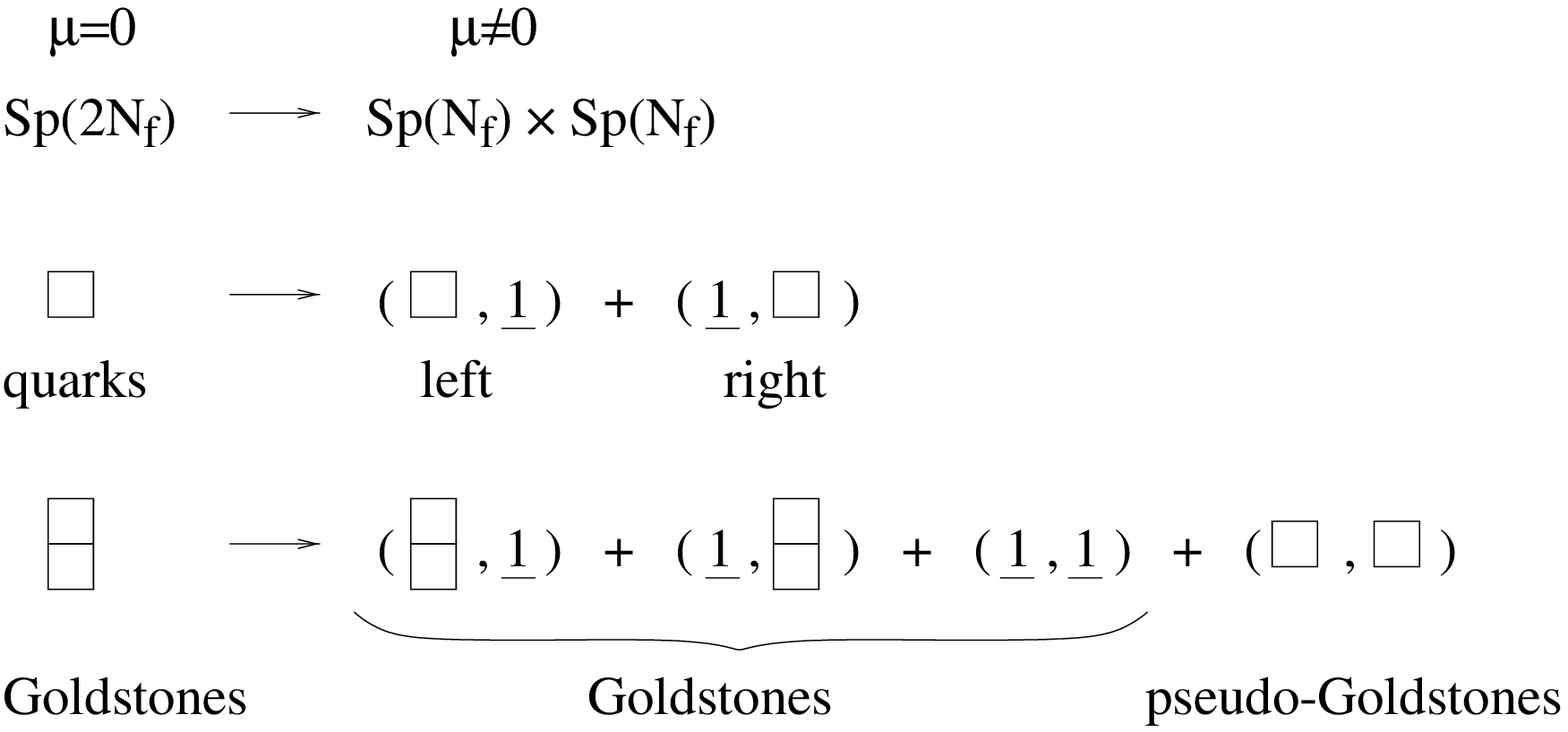,width=36em}}
\caption[]{Young's diagrams illustrating the breakdown of
the Goldstone fields into multiplets at $\mu=0$ and at $\mu\ne0$.
The quarks, which transform in the fundamental representation,
are given first as an example.
}
\label{fig:young}
\end{figure}

After the spontaneous breakdown (\ref{break1})
the $N_f^2$ pseudo-Goldstones are
degenerate and form an $(\underline{N_f},\underline{N_f})$
irreducible representation of the remaining manifest 
Sp($N_f$)$\times$Sp($N_f$).
The true Goldstones fall into 3 irreducible representations:
a singlet $(\underline1,\underline1)$, $(\underline{N_f(N_f-1)/2-1},1)$ and
$(1,\underline{N_f(N_f-1)/2-1})$, with the total count given by (\ref{gmune0}).
The irreducible representation 
$\underline{N_f(N_f-1)/2-1}$ of Sp($N_f$) is the antisymmetric 
tensor of rank 2 with the condition that the trace of that tensor times a
certain antisymmetric matrix vanishes (this representation does not
exist for $N_f=2$). This breakdown is convenient to
view in terms of Young diagrams in Fig.~\ref{fig:young}.

Understanding the multiplet structure turns out to be
very important in the analysis of the spectrum at small $\mu$ {\em
and} small quark mass $m_q$. This analysis will be presented
elsewhere.

\section{Conclusions}

In this letter we used two methods to study the physics of
2-color QCD at finite baryon chemical potential. We used
the fact that the measure of the Euclidean path integral in such
a theory remains positive definite even at finite $\mu$
to derive certain inequalities between non-singlet meson
correlators. These inequalities translate into inequalities
between masses of the lightest mesons and impose strong
restrictions on possible patterns of the symmetry breaking.
In particular, we show that the lightest meson is the $0^+$
diquark, and therefore condensation (if it occurs) must occur in the
channel $\psi^TC\gamma_5\psi$ thus leading to baryon charge 
superconductivity. This fact is in perfect agreement with
model calculations which show that both instanton-induced
and one-gluon exchange interactions are most attractive in this 
channel~\cite{SC}.

We also derived the low-energy effective Lagrangian describing
the mesons and baryons of the 2-color QCD with {\em massless}
quarks. We found that both the sign and the magnitude of 
the coefficient of the potential term of this
Lagrangian is fixed by a {\em local} chiral symmetry.
The sign determines the pattern of the spontaneous symmetry breaking,
and is such that it agrees with the QCD inequalities.
The masses of the mesons as a function of $\mu$
can be also determined exactly for small $\mu$. For example,
in the case of $N_f=2$ flavors of quarks the low energy
spectrum at small $\mu$ consists of one massless particle,
and a 4-plet of massive particles with masses equal to $2\mu$.

This result can be understood physically. The massless particle is the
Goldstone of the broken symmetry generated by the baryon charge. It
has the quantum numbers of a scalar diquark $\psi^TC\gamma_5\psi$. The 
SU(2)$\times$SU(2) 4-plet of massive mesons is comprised of the
sigma ($\bar\psi\psi$) and 3
pions ($\bar\psi\gamma_5\tau_i\psi$).  These excitations
could be thought of as loosely bound pairs, or threshold states of a
quark and an antiquark similar to the sigma particle in the
Nambu-Jona-Lasinio model~\cite{NJL} with mass $2m_{\rm fermion}$. In
the rest frame, the quark is taken from the surface of the Fermi sea,
with momentum $|\mbox{\boldmath $p$}|=\mu$, while the antiquark has
the opposite momentum $-\mbox{\boldmath $p$}$, thus making up the
invariant mass of $2\mu$. One should be aware, however, of the fact
that such a description is intuitive at best, since the ground state
can be described by the Fermi sphere only in the absence of
interactions between quarks. In the theory we consider the quarks are
confined.

We would like to thank S. Nussinov, E. Shuryak, J. Verbaarschot and
A. Zhitnitsky for very fruitful and stimulating discussions. Many of the
issues discussed here have been addressed in the context of Lattice Gauge
Theory. J.B.K. thanks M.-P. Lombardo, S. Hands, S. Morrison and D.K. Sinclair
for discussions. J.B.K. is supported in part by the National Science 
Foundation, NSF-PHY96-05199. M.S. is supported by the grant NSF-PHY97-22101.
D.T. is partially supported by the US DOE grant DE-FG-88ER40388 and
by ``Holderbank''-Stiftung.

\end{document}